\def\bC{{\beta_{_{C}}}}
\def\nL{{\mathcal{L}}}
\def\nW{{\mathcal{W}}}

\documentclass[materials,article,submit,pdftex,moreauthors]{Definitions/mdpi} 
\renewcommand{\linenumbers}{}
\firstpage{1} 
\pubvolume{1}
\issuenum{1}
\articlenumber{0}
\pubyear{2023}
\copyrightyear{2023}
\datereceived{ } 
\daterevised{ } % Comment out if no revised date
\dateaccepted{ } 
\datepublished{ } 
\hreflink{https://doi.org/} % If needed use \linebreak
\Title{Axion field influence on Josephson junction quasipotential}

\TitleCitation{Axion field influence on Josephson junction quasipotential}

\Author{Roberto Grimaudo $^{1}$\orcidA{}, Davide Valenti $^{1}$\orcidB{}, Bernardo Spagnolo $^{1,2}$\orcidC{}, Antonio Troisi $^{3}$\orcidD{}, Giovanni Filatrella $^{3,4}$\orcidE{} and Claudio Guarcello $^{5,4}$\orcidF{}}

\AuthorNames{Roberto Grimaudo, Davide Valenti, Bernardo Spagnolo, Antonio Troisi, Giovanni Filatrella and Claudio Guarcello}

\AuthorCitation{Grimaudo R.; Valenti D.; Spagnolo B.; Troisi A.; Filatrella G.; Guarcello C.}
\address{%
$^{1}$ \quad Dipartimento di Fisica e Chimica ``E. Segr\`e'', Group of Theoretical Interdisciplinary Physics, Universit\`a degli Studi di Palermo, Viale delle Scienze, Ed. 18, I-90128 Palermo, Italy;\\
$^{2}$ \quad Lobachevskii University of Nizhnii Novgorod, 23 Gagarin Ave. Nizhnii Novgorod 603950 Russia;\\
$^{3}$ \quad Dep. of Sciences and Technologies, University of Sannio, Via De Sanctis, Benevento I-82100, Italy;\\
$^{4}$ \quad INFN, Sezione di Napoli Gruppo Collegato di Salerno, Complesso Universitario di Monte S. Angelo, I-80126 Napoli, Italy;\\
$^{5}$ \quad Dipartimento di Fisica ``E.R. Caianiello'', Universit\`a di Salerno, Via Giovanni Paolo II, 132, I-84084 Fisciano (SA), Italy;}

\corres{Correspondence: cguarcello@unisa.it (C.G) giovanni.filatrella@unisannio.it (G.F.)}

\abstract{The direct effect of an axion field on Josephson junctions is analyzed through the consequences on the effective potential barrier that prevents the junction from switching from the superconducting to the finite-voltage state. 
We describe a method to reliably compute the quasipotential with stochastic simulations, which allows to span the coupling parameter from weakly interacting axion to tight interactions.
As a result, we obtain that the axion field induces a change in the potential barrier, therefore determining a significant detectable effect for such a kind of elusive particle. }


\begin{document}

%%%%%%%%%%%%%%%%%%%%%%%%%%%%%%%%%%%%%%%%%%
\section{Introduction}

Nowadays, in the search for cold dark matter candidates, among others, axion particles were theoretically predicted, but their detection remains elusive, for the very weak interaction that they are supposed to have with ordinary matter~\cite{Preskili83,Abb83,Din83}.
Since Josephson junctions (JJs) proved to be very sensitive devices, close to detecting a single photon~\cite{Alesini20}, a natural idea was to exploit them to detect the electromagnetic field produced by an axion decay~\cite{Rettaroli21}. In this case, the detection of primordial axions results from their decay into microwave photons in a resonant cavity through the inverse Primakoff effect at rather high magnetic fields ($\sim3-8$ T), which is quite feasible using superconducting magnets, assuming also a quantum-limited parametric amplifier or single photon detector, placed outside the solenoid and inside magnetic shields~\cite{Barbieri17,McAllister17,Crescini20}.
A different possibility, which is the framework for the present paper, is to exploit the direct interaction between JJ and axions~\cite{Bec13} in a detector~\cite{Grimaudo22,Gri23}. This scheme was also proposed as a key to understand unclear ``events'' in Josephson's response~\cite{Bec17,Hof04,Bae08,He11,Gol12,Bre13,Wang22} and would simplify the usual detection schemes through a reduced setting.
To date, the main idea behind this direct interaction is that the axion decays inside the junction with a rather high probability, which would be achieved in a resonant cavity through the Primakoff effect only through a huge magnetic field (orders of magnitude above any realistic field): thus, a ``Josephson cavity'' is much more effective to detect the axion than a resonant cavity~\cite{Bec13} .
However, this scheme misses a full theory of the Josephson-axion interaction and a detailed description of the changes induced in the JJ dynamics, which could possibly lead to detectable consequences. The idea behind the original proposal hides a shortcoming, namely the fact that the actual axion-JJ coupling mechanism is not well substantiated, and thus a sensitivity of the method cannot be defined.
Thus, we concentrate here only on the nonlinear dynamic problem, assuming that the axion-JJ interaction exists, although the interaction parameter is unknown.
In the first proposal it was suggested to look at deviations from the locked dynamics of the JJ to an external radio frequency -- the so-called Shapiro steps. 
More recently, it has been proposed by some of the authors of the present paper to assume a different standpoint: to bias the JJ in the superconducting metastable state through a dc external drive and to observe the passages to the finite voltage state in the presence of the axion-JJ coupling, under the influence of thermal fluctuations~\cite{Grimaudo22}. 
The idea is that these switches are altered by the interaction of the JJ with the axions, and it is therefore possible to infer the existence of the axions if the switching is, in some statistical sense, changed; a more detailed analysis can also offer an estimate of the JJ-axion coupling from switching time measurements~\cite{Grimaudo22}. A successive approach assumed the JJ operating as a qubit, and in such a way the qubit-axion interaction being detected as axion-induced oscillations of the qubit state~\cite{Gri23}.
The general idea of exploiting a JJ to detect a weak signal, even embedded in a noisy background, is fairly well-established, a JJ being essentially a threshold device operating via a switching mechanism~\cite{Braginski2019,Tafuri2019}. The presence of a noise background is a condition typical of complex systems, such as, for instance, ecological systems~\cite{Valenti16} and financial markets~\cite{Valenti18}, which has to be taken into account in view of better modeling their dynamics. Josephson junctions have been also proposed as noise detectors~\cite{Tob04,Pek04,Ank07,Suk07,Tim07,Hua07,Gra08,Fil10,Gua13,Gua20,Gua21-2} and play a leading role in the search for possible protocols and schemes for the detection of single photons~\cite{Wal17,Kuz18,GuaBra19,Rev20,Yab21,Pied21,Gua21,Pan22-1,Pan22-2}. 

In this paper we further investigate the consequences of a direct interaction between axions and JJ, to show that a certain quantity, namely, the \emph{quasipotential}~\cite{Graham85}, can be introduced for this non-equilibrium system and that it is possible to determine the quasipotential barrier in the presence of the axions.
To investigate the quasipotential is an advantage, as it can be determined with numerical simulations at a relatively high noise intensity, i.e., high temperature. 
Indeed, the quasipotential (as the ordinary potential) is not affected by noise; if the quasipotential is known, it is possible to predict the average escape time at very low temperature with the Arrhenius law, with a considerable saving of simulation time~\cite{Kau88}. 
We do so with a twofold objective: in the first place, as already mentioned, to better understand the consequences that a direct interaction between JJ and axion would have, and therefore to pave the way towards a practical implementation of the device to detect axions; on the second hand, not less important, to compute the quasipotential barrier of the axion-JJ system, a very convenient quantity in the analysis of low temperature devices (as already demonstrated for Shapiro steps~\cite{Kautz1995,Kau96}, cavity-induced synchronization~\cite{Pou19}, the study of a JJ electronic analogue~\cite{McConnell2021}, and also in non-Josephson contexts ~\cite{Dykman1998,Kraut2003,Kraut2003b,Khovanov2014,Li2019,Frimmer2019}) to infer the properties at a very low noise and, consequently, very long escape times. In fact, the characteristic time scale of the system is of the order of $[1-10]\;\text{ps}$, being the inverse of Josephson characteristic frequency which, as we shall see later, generally falls in the range $[0.1-1]\;\text{THz}$; this means that numerical realizations, even close to real experimental times (which could take up to milliseconds, e.g., for experiments involving switching current distributions), require extremely long simulations. Moreover, these have to be repeated several times in order to obtain complete statistics. Indeed, stochastic analyses, such as the one we propose, require the repetition of the same experiment, i.e., of the same numerical simulation, for a reasonably large number of times under the same conditions, in order to allow for reliable statistical analyses. In conclusion, quasipotential analysis makes it possible to extract useful information at reasonably high temperatures, that means within reasonable simulation times, and then allows to extrapolate relevant information even at low temperatures, where numerical simulations time would become prohibitive. 

The paper is organized as follows: Sect.~\ref{Model} presents the model to describe: the JJ (Sect.~\ref{ModelJJ}), the axion field (Sect.~\ref{Modelaxions}), and the interacting axion-JJ system (Sect.~\ref{ModelJJaxions}). Sect.~\ref{Results} defines the quasipotential for this system and computes its behavior as a function of the interaction. 
Finally, in Sect.~\ref{Conclusions} conclusions are drawn.

\section{Model} 
\label{Model}

In this section we outline the models for the JJ, see Sect.~\ref{ModelJJ}, the axion, see Sect.~\ref{Modelaxions}, and their interaction, see Sect.~\ref{ModelJJaxions}. 
We show the potential of the JJ alone -- a cosinusoidal washboard potential, see Eq.~\eqref{Washboard App} below -- that exhibits an activation energy barrier, whose changes due to the interaction with the axions are the focus of the present work.
Also, some details for the numerical simulations of the stochastic equations are given in Sect.~\ref{ModelJJ}.

\subsection{RCSJ Model}
\label{ModelJJ}

\begin{figure}[t!!]
\centering
\includegraphics[width=0.7\textwidth]{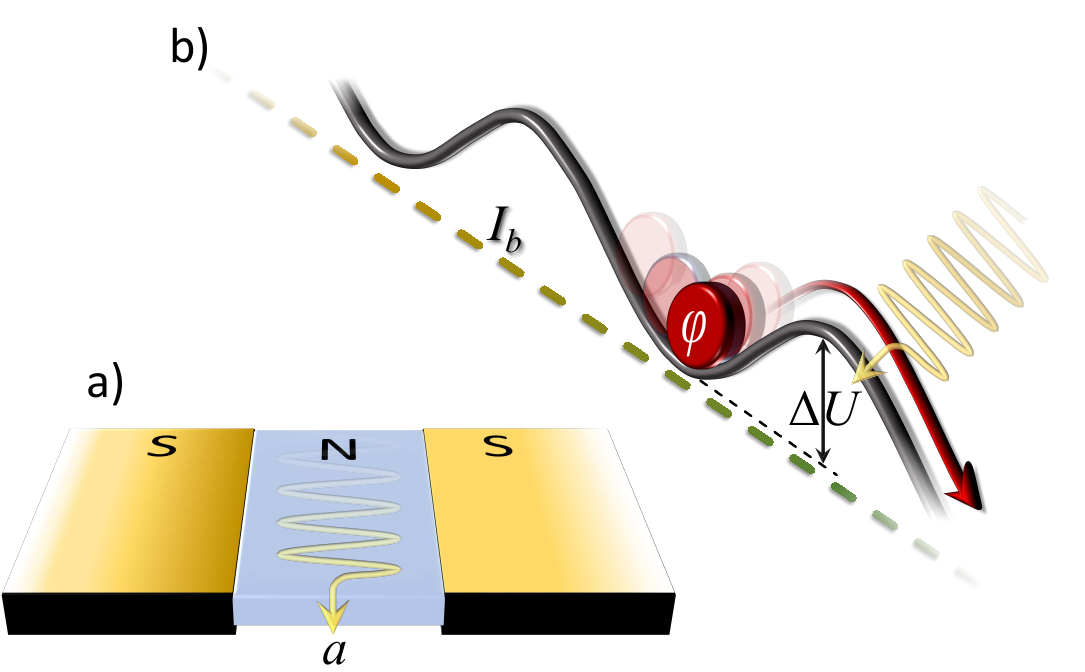}
\caption{\label{fig: Device} Schematic illustration of the device. (a) The physical process: An axion field $a$ enters the normal barrier between two superconducting electrodes. (b) The mathematical model: The phase particle in a minimum of the washboard potential $U$, tilted by a bias current, is perturbed by the axion (represented by the arrow that hits the particle-phase $\varphi$). Under the combined effect of thermal noise and axion-JJ coupling, the JJ phase can overcome the barrier $\Delta U$ and roll down along the potential. In this case, a detectable voltage $\propto \langle d\varphi/dt\rangle $ appears.}
\end{figure}

Let us consider the usual model for a superconducting junction, schematically represented in Fig.~\ref{fig: Device}(a), given by the following equations~\cite{Bar82,Lik86}
\begin{eqnarray}
\label{JJcurrent}
I_\varphi = I_c \sin{\varphi},\\
\label{JJvoltage}
V = \frac{\Phi_0}{2\pi}\frac{d\varphi}{dt},
\end{eqnarray}
where $\Phi_0=h/(2e)$ is the flux quantum, with $e$ and $h$ being the electron charge and the Planck constant, respectively, $I_c$ is the maximum Josephson current that can flow through the device, and $\varphi$ is the Josephson phase difference.

For a real device, one assumes for instance that the two superconductors have lateral dimensions $\nL$ and $\nW$ smaller than the Josephson penetration depth, $\lambda_{_{J}} = \sqrt{\Phi_0/(2\pi \mu_0 t_d J_c)}$ (here, $t_d=\lambda_{L,1}+\lambda_{L,2}+d$ is the effective magnetic thickness, with $\lambda_{L}$ and $d$ being the London penetration depths and the insulating layer thickness, respectively, $\mu_0$ is the vacuum permeability, and $J_c$ is the critical current area density).
The dynamics of the Josephson phase $\varphi$ for a dissipative, current-biased small JJ can thus be studied within the resistively and capacitively shunted junction (RCSJ) framework~\cite{Bar82,McC68,Gua19,Gua20}

\begin{equation}
\left ( \frac{\Phi_0}{2\pi} \right )^{\!\!2}\!\! C \frac{d^2 \varphi}{d t^2}+\left ( \frac{\Phi_0}{2\pi} \right )^{\!\!2}\!\!\frac{1}{R} \frac{d \varphi}{d t}+\frac{d }{d \varphi}U 
= \left ( \frac{\Phi_0}{2\pi} \right )( I_N+I_b),
\label{RCSJ App}
\end{equation}
with $R$ and $C$ the normal-state resistance and capacitance of the JJ, respectively, and $I_N$ and $I_b$ the thermal noise and the bias current, respectively.
The corresponding normalized dynamics can be reformulated (for sinusoidal potential of standard tunnel JJ, albeit other shapes are possible~\cite{Bee92}) through the equation
\begin{equation}
\beta_c\frac{d^2 \varphi (\tau_c)}{d\tau_c^2}+ \frac{d \varphi (\tau_c)}{d\tau_c} +\frac{d }{d \varphi}\mathcal{U}(\varphi,i_b) = i_{n}(\tau_c) + i_b,
\label{RCSJnormOc}
\end{equation}
where time is normalized to the inverse of the characteristic frequency, that is $\tau_c = \omega_c~t$ with $\omega_c=\left ( 2\pi/\Phi_0 \right )I_cR$, $i_b= I_b / I_c$ and $i_n= I_n / I_c$ are the normalized external bias current and thermal noise current, and $\beta_c=\omega_c RC$ is the Stewart-McCumber parameter. 
We stress that the JJ response is usually quite fast, since the characteristic frequency of JJ falls within the range $ [0.1,1]\;\text{THz}$. 
Another way to obtain a dimensionless form of Eq.~\eqref{RCSJ App} consists in normalizing with respect to the plasma frequency $\omega_p=\sqrt{2eI_c/\hbar C}$.
In this case, time is normalized respect to the inverse plasma frequency, i.e., $\tau_p = \omega_p~t$, and the equation in normalized units contains a damping parameter $\alpha=\beta_{_{C}}^{-1/2}$, which multiplies the first time-derivative of the phase. 

The normalized potential, $\mathcal{U}$, is the so-called \emph{washboard potential}, which depends upon the normalized bias current, $i_b$, and the Josephson phase according to
\begin{equation}
\mathcal{U}(\varphi,i_b)=\frac{U(\varphi,i_b)}{{E_{J_0}}}=\left [1- \cos(\varphi) -i_b\varphi\right ].
\label{Washboard App}
\end{equation}
The potential can be expressed in physical units defining the Josephson energy $E_{J_0}=\left ( \Phi_0/2\pi \right )I_c$. 
The resulting activation energy barrier, $\Delta U(i_b)$, confines the phase $\varphi$ in a metastable potential minimum and can be calculated as the difference between the maximum and minimum value of the normalized potential $U(\varphi,i_b)$, see Fig.~\ref{fig: Device}(b).
In units of $E_{J_0}$, it can be expressed as
\begin{equation}
{\Delta \mathcal{U}(i_b)}=\frac{\Delta {U}(i_b)}{{E_{J_0}}}=2 \left [ \sqrt{1-i_b^2} -i_b\arccos(i_b)\right ].
\label{activationenergybarrier App}
\end{equation}
In the phase particle picture, the term $i_b$ represents the tilting of the potential profile; increasing $i_b$ the slope of the washboard increases and the height 
$\Delta \mathcal{U}(i_b)$ of the rightward potential barrier reduces, until this activation energy vanishes altogether for $i_b=1$, that is when the bias current reaches its critical value $I_c$. 
During the motion, different regimes are governed by the Stewart-McCumber parameter $\bC$.
 A highly damped (or overdamped) junction corresponds to $\bC\ll 1$, that is a small capacitance and/or a small resistance. 
 Instead, a junction with $\bC\gg 1$ has a large capacitance and/or a large resistance, and is weakly damped (or underdamped)~\footnote{With the alternative normalized mentioned before, the under- and overdamped regimes correspond to $\alpha \ll 1$ and $\alpha \gg 1$, respectively.}.
 For the purposes of this work, it is important to notice that in the underdamped regime once the phase has passed the barrier, a finite velocity, and hence, as per Eq.~\eqref{JJvoltage}, a finite voltage, appears.
It is therefore possible to detect the passage of the Josephson phase over the barrier through the appearance of a finite voltage, a key point to employ a JJ as a detector. In fact, the phase $\varphi$ itself is not directly accessible, while the passage over the barrier is signaled by a measurable voltage drop across the junction.
The procedure can be briefly schematized as follows. 
The JJ is prepared in some static configuration (at which corresponds a zero net voltage), exposed to some supposedly existing perturbation, and the junction is left to evolve. 
If the signal was not present, the passage only occurs under the effect of thermal noise, and it is given by the usual Kramers law~\cite{Kra40}. The presence of the signal is ascertained through deviations of the thermal escapes~\cite{Fil10,Pied21} -- as will be discussed in more details below.

In this work, the random current is modeled as a delta-correlated Gaussian white noise associated to the normal-state resistance of the junction, $R$, with the usual statistical properties
\begin{eqnarray}
\label{averageD}
\langle i_n (\tau)\rangle\ &=& 0,\\
\label{sigmaD}
\langle i_n (\tau) i_n (\tau+\tilde{\tau})\rangle &=& 2D\,\delta (\tilde{\tau}). 
\end{eqnarray}
The amplitude of the normalized correlation is connected with the physical temperature $T$ through the relation~\cite{Bar82}
\begin{eqnarray}
\label{WNAmp}
D= \frac{k_BT}{R}\frac{\omega_c}{I^2_c},%=\frac{k_BT}{E_{J_0}},
\end{eqnarray}
here $k_B$ is the Boltzmann constant.
We note that, by normalizing time with respect to the characteristic frequency $\omega_c$ (as we do in our numerical simulations), the normalized noise intensity in Eq.~\eqref{WNAmp} can be recast as $D={k_BT}/{E_{J_0}}$, i.e, the ratio between the thermal energy and the Josephson coupling energy, $E_{J_0}$, without reference to the damping; instead, normalizing with respect to the plasma frequency, $\omega_p$, the normalized noise intensity becomes $D=\alpha{k_BT}/{E_{J_0}}$.
Thus, for Gaussian fluctuations of amplitude $D$, the stochastic independent increment employed in the numerical simulations reads
$\Delta i_N \simeq \sqrt{ 2 D \Delta t\; }\; N\left(0,1 \right)$.
Here, $N\left(0, 1 \right)$ indicates a Gaussianly distributed random function with zero mean and unit standard deviation.

%Another way to obtain a dimensionless form of Eq.~\eqref{RCSJ App} consists in normalizing with respect to the plasma frequency $\omega_p=\sqrt{2eI_c/\hbar C}$.
%In the latter case, the normalized RCSJ equation \eqref{RCSJ App} reads
%
%
%\begin{equation}
%\frac{d^2 \varphi (\tau_p)}{d\tau_p^2}+ \alpha \frac{d \varphi (\tau_p)}{d\tau_p} + \sin \left [ \varphi\left ( \tau_p \right ) \right ] = i_{n}(\tau_p) + i_b,
%\label{RCSJnormOp App}
%\end{equation}
%
%
%where time is normalized respect to the inverse plasma frequency (that reads $\omega_p=\sqrt{{\Phi_0}/{2\pi C}}$ ), $\tau_p = \omega_p~t$, $\alpha=1/\sqrt(\omega_p~R~C)=1/\sqrt{\bC}$ is the damping parameter. 
%With this time normalization the under- and over-damped regimes correspond to $\alpha \ll 1$ and $\alpha \gg 1$, respectively.

\subsection{Axion}
\label{Modelaxions}

If one describes the axion field $a$ in the Friedman-Robertson-Walker metric, the equation of motion of the axion misalignment angle $\theta$ and the axion coupling constant $f_a$, namely $a=f_a\,\theta$~\cite{Sik83,Vis13}, reads
\begin{equation}
\frac{d^2 \theta (t)}{dt^2}+ H \frac{d \theta (t)}{dt} + \frac{m_a^2c^4}{\hbar^2} \sin \left [ \theta \left ( t \right ) \right ] = 0,
\label{AxionEq}
\end{equation}
complemented with spatial gradients that are here omitted. 
The above model includes the forcing term $\sin(\theta)$ due to quantum chromodynamics instanton effects. As one can observe, there is a formal similarity between the equation of motion governing the axion and the RCSJ systems, being the axion dynamics analogous to an unbiased RCSJ. Moreover, in normalized units, the parameters are of the same order of magnitude.
In Eq.~\eqref{AxionEq}, $H \approx 2 \times 10^{-18} ~ s^{-1}$ is the Hubble parameter and $m_a$ is the axion mass. 
The typical ranges of parameters that are allowed for dark matter axions are~\cite{sik09,Duf09}: 
$ 3 \times 10^9 ~ \text{GeV} \leq f_a \leq 10^{12} ~ \text{GeV}$ and $ 6 \times 10^{-6} ~ \text{eV} \leq m_a c^2 \leq 2 \times 10^{-3} ~ \text{eV}$. 
The prediction of the axion's mass, based on the average of the results from five independent condensed matter experiments, is $ m_a c^2 = (106 \pm 6) \mu eV$~\cite{,Hof04,Bae08,He11,Gol12,Bre13,Wang22}.

\subsection{Axion-JJ System}
\label{ModelJJaxions}

According to the approach of Refs.~\cite{Grimaudo22,Yan20}, the interaction between axion and JJ can be formally written as 
\begin{subequations}\label{Orig Diff Eqs Syst}
\begin{align}
\ddot{\varphi} + a_1 \dot{\varphi} + b_1 \sin(\varphi) &= \gamma (\ddot{\theta} - \ddot{\varphi}) \label{Orig Diff Eqs Syst a},\\
\ddot{\theta} + a_2 \dot{\theta} + b_2 \sin(\theta) &= \gamma (\ddot{\varphi} - \ddot{\theta}),
\end{align}
\label{Orig Diff Eqs Syst}
\end{subequations}
\noindent where $(a_1, a_2)$ and $(b_1, b_2)$ are the dissipation and frequency parameters, respectively; $\gamma$ is the coupling constant between the two systems, whose values one wants to infer from the experiments. 
This model, which succeeds in explaining some experimental anomalies~\cite{Hof04,Bae08,He11,Gol12,Bre13,Wang22}, is based on the possibility to formally treat the axion as an effective JJ, and therefore to consider the system in Eqs.~\eqref{Orig Diff Eqs Syst} as equivalent to two capacitively coupled JJs~\cite{Gordeeva06,Blac09}.

To model the Josephson phase dynamics with a bias current and thermal fluctuations, Eqs.~\eqref{Orig Diff Eqs Syst} can be conveniently rewritten as (see details in Ref.~\cite{Grimaudo22} -- Appendix B)
\begin{subequations}
\label{Diff Eqs Syst omegac}
\begin{align}
%\label{DiffEqsJJ}
{\beta_c \over k_2}~\ddot{\varphi}+\dot{\varphi}+\sin(\varphi)+{k_1 \over k_2}~\varepsilon~\sin(\theta) &= i_b+i_n, \label{Diff Eqs Syst omegac phi} \\
%\label{DiffEqsaxion}
{\beta_c \over k_1}~\ddot{\theta}+\dot{\varphi}+\sin(\varphi)+{k_2 \over k_1}~\varepsilon~\sin(\theta) &= i_b+i_n, \label{Diff Eqs Syst omegac theta}
\end{align}
\end{subequations}
with 
\begin{equation}
\begin{aligned}
&k_1  =  {\gamma \over 1+2\gamma}, \qquad k_2  =  {1+\gamma \over 1+2\gamma}, \qquad \text{and}\qquad \varepsilon  =  \left(\frac{m_ac^2}{\hbar\omega_p}\right)^2.
\label{epsilon}
\end{aligned}
\end{equation}
%
%&k_1 = {\gamma \over 1+2\gamma}, \qquad k_2 = {1+\gamma \over 1+2\gamma}, \\
%& \beta_c = \left( \frac{\omega_c}{\omega_p} \right)^2,\quad \varepsilon = \left({m_ac^2 \over \hbar\omega_p}\right)^2,
%
%
%
 The $\varepsilon$ parameter indicates the ratio between the axion energy and the 
Josephson plasma energy, $\hbar\omega_p$, and can be chosen -- within the JJ fabrication constraints -- to select the most convenient working point for the detection of an axion field interacting with the JJ. 
Indeed, the Josephson plasma frequency, and therefore the energy ratio $\varepsilon$, can be ``adjusted'' as needed, because $I_c$ can be lowered either by raising the temperature~\cite{Dub01} or by applying a magnetic field~\cite{Ber08} or a gate voltage~\cite{Du08}. 
The purpose is to determine the working point at which the system is most responsive to the axion perturbation. 
As the detection is performed through the analysis of the escape times, the response is measured in the precise sense that the distribution of the escape times for the axion-JJ coupled system deviates the most from the Josephson response in the absence of axions. In Ref.~\cite{Grimaudo22} it was in fact showed that at $\varepsilon\lesssim1$ the average switching time approaches a minimum due to an axion-induced resonant activation phenomenon, for the occurrence of an effective frequency matching between axion and JJ, and also that the optimal experimental conditions for a JJ-based axion detection scheme should involve a Josephson plasma energy lower than the axion energy, i.e., $\varepsilon>1$.

In this work, we trace the change in the escape time (that makes the axion-JJ interaction detectable) back to the change in the effective potential barrier that confines the system to the static zero-voltage configuration. In the following Sect.~\ref{Results} we discuss how to compute the effective energy.

The integration of the stochastic Eqs.~\eqref{Diff Eqs Syst omegac} is performed with a finite-difference explicit method, the partial derivative are approximated using the Euler formalism with an integration time integration step $\Delta t=10^{-2}$, a maximum integration time $t_{max}=10^6$, initial conditions $\varphi(0)=\arcsin{(i_b)}$  and $\theta(0)=\dot{\varphi}(0)=\dot{\theta}(0)=0$, and repeating each simulation $N=10^4$ times for each set of parameter values. Other parameters useful for the calculations are $\beta_c=100$ (i.e., underdamped regime) and $\varepsilon=1$.

\section{Calculation of the Quasipotential}
\label{Results}

\begin{figure}[t!!]
\begin{center}
\includegraphics[width=\columnwidth]{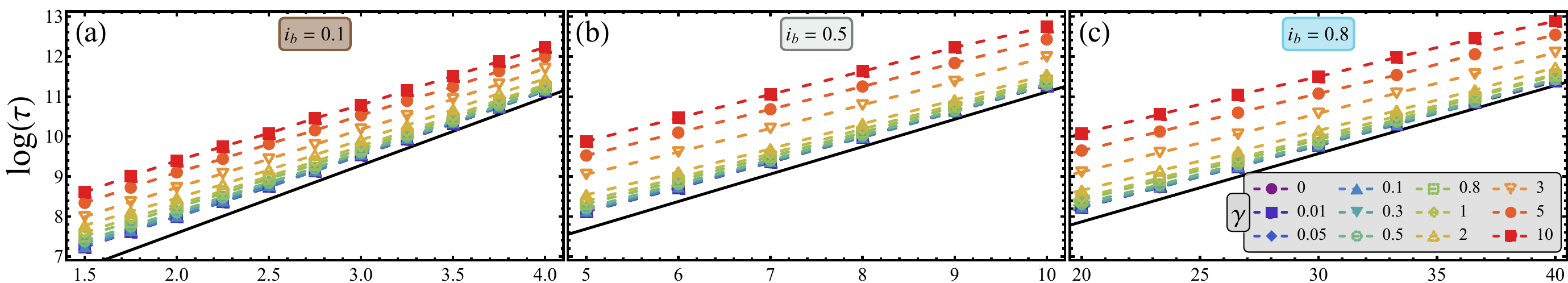}
\includegraphics[width=\columnwidth]{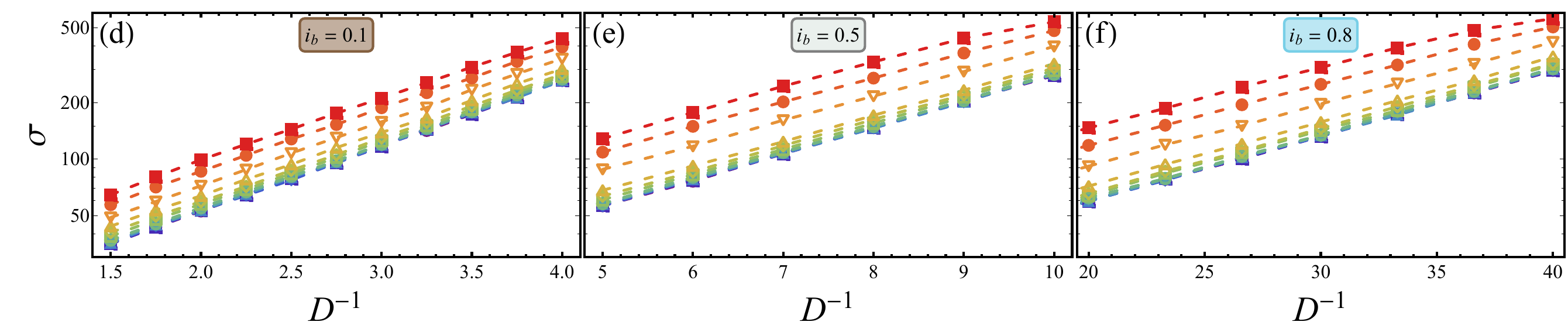}
%\captionsetup{justification=raggedright,format=plain,skip=4pt} 
\caption{\label{Kramers_fit} Behavior of the average value (top panels) and the root-mean-square (bottom panels) of the switching time distributions as a function of the inverse noise intensity, for different values of the axion-JJ coupling, $\gamma$, and three different bias current, $i_b=\{0.1, 0.5,\text{ and }0.8\}$, see panels (a, d), (b, e), and (c, f), respectively. The black solid line without symbols in the top panels denotes the full Kramers theory for the dissipative Josephson escape~\cite{Risken89}. The other parameters of the system are $\varepsilon=1$, $\beta_c=100$, $t_{max}=10^6$, and $N=10^4$. The legend in panel (c) refers to all panels.}
\end{center}
\end{figure}

The non-equilibrium system in Eq.~\eqref{Orig Diff Eqs Syst} does not admit an ordinary potential. However, it is possible to define an effective, or quasi, potential that keeps the system in the static configuration. The axion-JJ coupled system eventually switches from the superconducting state ($V\propto d\varphi/dt=0$) to the resistive state ($V\propto d\varphi/dt\neq 0$), when the combined effect of noise and axion interaction allows the JJ to overcome the effective energy barrier. 
In this picture, one can think of the axion effect on the JJ as some perturbation that changes (more precisely, lowers, as we shall demonstrate below) the effective energy barrier. The advantage is that the change of this effective energy is independent of noise and therefore holds at any (sufficiently low) temperature.
The main difficulty is therefore to determine how the coupling between Eqs.~\eqref{Diff Eqs Syst omegac} amounts to a change in the quasipotential barrier. 
To begin with, we show how to compute the quasipotential barrier, that is the effective energy that must be overcome to induce a switch. 
The basic logic is as follows: suppose that the Arrhenius behavior~\cite{Graham85}
\begin{equation}
\label{Kramers}
\tau = \lim_{D\to0}\tau_0 \exp{\frac{\kappa}{D} }
\end{equation}
is valid. 
The hypothesis obviously holds for the "pure" JJ system, i.e, Eq.~\eqref{RCSJnormOc}, for which $\kappa\equiv\Delta \mathcal{U}$. 
One can make the further conjecture that the average escape time is exponential in the inverse of the noise intensity, even in the $\kappa \neq \Delta \mathcal{U}$ case, such that, in the limit of small noise,
\begin{equation}
\label{exprelation}
 \log{\frac{\tau}{\tau_0}} = \kappa \frac{1}{D}.
\end{equation}
Under general assumptions~\footnote{For out-of-equilibrium systems that do not admit an ordinary potential, it is possible to define a non-equilibrium potential with properties analogous to those of an ordinary potential, as long as there is a single time-independent probability distribution that can be reached from any initial distribution as the weakly stochastic dynamical system approaches its steady state~\cite{Graham86}.}, it is fair to interpret the coefficient $\kappa$ as an effective energy barrier
\begin{equation}\label{quasipotential}
\Delta \mathcal{U}_{eff} \equiv \kappa = \lim_{D^{-1} \rightarrow \infty}\frac{ \log{\tau/\tau_0}}{D^{-1}}.
\end{equation}
In other words, the slope $\kappa$ of the relation \eqref{exprelation} can be interpreted as a {\it bona fide} potential barrier, in the limit of small noise. 
The advantage of this interpretation is twofold. 
On the one end, it gives a physically intuitive interpretation to the effect of the axion field; as we shall prove in the following subsection, the axion-JJ coupling $\gamma$ lowers the confining barrier. 
On the other hand the quasipotential offers a practical advantage, because it allows to extrapolate the results to very low values of noise, that is in the region where escape times are prohibitively long and extremely difficult to reach with simulations.
It is in fact enough to enter the regime of exponential decay to determine $\kappa$, and then to exploit such value for any lower value of the noise intensity $D$. 
The effective potential~\eqref{quasipotential} can be numerically retrieved with several estimates of the escape time $\tau$ as a function of the noise amplitude $D$; in the plot $\log{\tau}$ vs $1/D$ the prefactor $\tau_0$ is the $y$-axis intercept and $\kappa$ the slope of the relationship. 
More precisely, the two quantities should be computed, for different values of the axion-JJ coupling $\gamma$ and the bias current $i_b$, in the exponential regime, that is discarding the data for high $D$. This ensures that the asymptotic regime \eqref{Kramers} has been entered, as shown in the top panels of Fig.~\ref{Kramers_fit}. The bottom panels of this figure serves to compare the averages with the root-mean-squares, $\sigma$, of the switching time distributions, as a function of the inverse of the noise amplitude, as is often done when studying the Josephson switching dynamics~\cite{Gordeeva06,Pankratov2012,Yablokov2021}.

%\subsection{Behavior of the Quasi-Potential}

The axions are revealed through the difference between the switching time distributions of a JJ without external perturbation and under the influence of an axion field. However, for any finite sampling the actual measured average is subject to fluctuations. Therefore, the best detection is obtained with the method to which pertains the best signal-to-noise ratio (SNR). Here, a possible signal is the measured average difference between the switching times with and without axion-induced perturbation. The noise is due to the fluctuations around the average switching time, i.e, the standard deviation of the sampled mean; the SNR thus computed is often measured through the Kumar-Carroll index~\cite{Kumar84}. However, for simplicity in this work we merely observe the effect of axion on the effective potential felt by the axion-JJ system.

\begin{figure}[t!]
\begin{center}
\includegraphics[width=\textwidth]{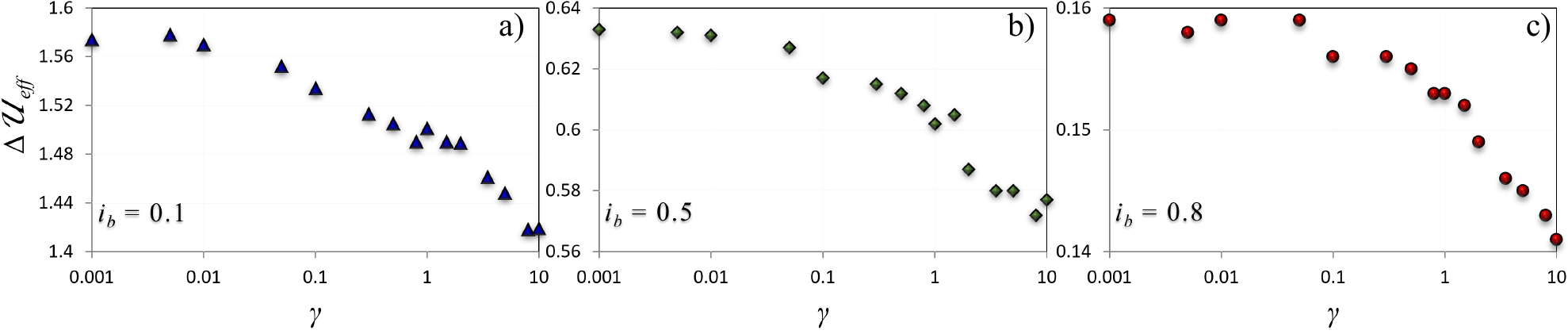} 
\captionsetup{justification=raggedright,format=plain,skip=4pt}
\caption{\label{fig:QP} Behavior of the quasipotential effective barrier, $\Delta \mathcal{U}_{eff}$, retrieved from the data shown in Fig.~\ref{Kramers_fit}, as a function of the axion-JJ coupling $\gamma$. Panels a), b), and c) refer to different bias current values, i.e., $i_b=\{0.1, 0.5,\text{ and }0.8\}$, respectively.}
\end{center}
\end{figure}

For the uncoupled regime, $\gamma =0$, the numerical quasipotential barrier should be equal to the washboard potential barrier. From Fig.~\ref{fig:QP} we can estimate a modest discrepancy, $\eta=\left ( \Delta \mathcal{U}-\Delta \mathcal{U}_{eff} \right )\Big/\Delta \mathcal{U}<10\%$, due to both the finite number of realizations over which the average is calculated and the finite temperature. In fact, the quasipotential barrier should be computed in the limit $D^{-1}\rightarrow \infty$, or more accurately ($\Delta \mathcal{U}_{eff}/D) \rightarrow \infty$. 
This discrepancy, obtained as the percentage difference between $\Delta \mathcal{U}_{eff}$ estimated from the data of Fig.~\ref{fig:QP} at the lowest coupling, i.e., $\gamma=0.001$, and the analytical washboard activation energy $\Delta \mathcal{U}$, see Eq.~\eqref{activationenergybarrier App}, reads $ \eta=\{7.4\%, 8.0\%,\text{ and }6.1\%\}$ for $i_b=\{0.1, 0.5,\text{ and }0.8\}$, respectively.
It is indeed remarkable that, despite the relatively high noise (i.e., we obtain $\Delta \mathcal{U}_{eff}/ D \in [3-6]$ from data in Figs.~\ref{Kramers_fit} and~\ref{fig:QP}), the agreement is good.
This observation highlights the advantage of the quasipotential method, because one can use not too long escape times to effectively evaluate the effective quasipotential barrier through Eq.~\eqref{quasipotential}. Shortly, we have demonstrated that the quasipotential for the JJ-axion system can be numerically evaluated with relative ease, while the result can be extrapolated to much lower values of noise, and hence to much longer escape times.

Finally, we want to exploit the estimation of the effect of the quasipotential for the detection of the axion. 
This is summarized in Fig.~\ref{fig:QP}, where $\Delta \mathcal{U}_{eff}$, i.e., the slope of the escape times in Fig.~\ref{Kramers_fit}, versus $\gamma$ is just the quasipotential barrier.
For each of the three different values of the bias current considered, $i_b=\{0.1, 0.5,\text{ and }0.8\}$, it is proven that the increase of the coupling $\gamma$ lowers the effective energy barrier. 
%The effect seems more uniform for $i_b = 0.1$ and $0.5$, while for shallow barriers [e.g., see $i_b=0.8$ in Fig.~\ref{fig:QP}c)] the change is more evident only for $\gamma\gtrsim0.1$. In fact, 
The greater $i_b$, the larger the $\gamma$ value above which the coupling with the axion produces an effect on the quasipotential barrier; specifically, for $i_b = 0.1$ ($i_b=0.8$) the change is more evident only for $\gamma\gtrsim0.01$ ($\gamma\gtrsim0.1$). Interestingly, the bias current, which actually represents an easily controllable parameter, was also demonstrated to have a significant impact on the emerging of resonances in the switching times discussed in Ref.~\cite{Grimaudo22}.
It is therefore evident from our simulations that a lower bias current is more convenient. Moreover, a different $\gamma$ gives quite different quasipotential barrier heights, such that the overall effect of the coupling between JJ and axion is to reduce the effective height of the potential barrier. 
%\newpage

\section{Conclusions}
\label{Conclusions}

It has been shown that if the direct interaction between a solid state superconducting device, i.e., a JJ, and the dark matter candidate named axion is assumed, a modification of the response to noise of the former arises \cite{Grimaudo22,Gri23}. This aspect offers an opportunity for the detection of this elusive particle.
Using JJs to detect axions has been shown to be beneficial for several reasons. First, JJs are superconducting devices that can operate at very low temperatures, and, hence, at very low noise. Second, they are very fast elements, with typical characteristic frequencies from GHz to THz, and therefore large amount of data can be collected in a brief time. Third, some parameters of the Josephson device can be adjusted to tune the coupling with the axion. The bias current is a further degree of freedom that can be exploited to tune the effective barrier, as shown in Fig.~\ref{fig:QP}.

In a nutshell, axion signature can be sketched as follows: the JJ-axion interaction facilitates, in the presence of noise, the escapes of the Josephson phase from the superconducting to the finite-voltage state. 
This change can be described through the quasipotential, which is an effective energy barrier that summarizes the response in the limit of small noise. 
The introduction of the quasipotential allows to extrapolate the behavior at very low noise values, at which numerical simulations become prohibitively long. 
It is thus possible to reconstruct the response at low temperature through simulations performed at relatively high noise intensities, an advantage that has been already exploited in several applications, as for instance the Josephson voltage standard~\cite{Clark2000} for which even very rare escapes are relevant to maintain the high accuracy required by metrological standards~\cite{Kau96}. 
Analogously, for weak signal detection it is important to mimic occurrence of rare events in a quite noisy environment~\cite{Pied21}.
In this work we have extended the method to the interaction between an underdamped JJ and an axion field. 
Within this framework, it has been possible to demonstrate that the interaction is summarized by a quasipotential, and to determine the behavior of this effect through the quasipotential barrier as a function of the JJ-axion interaction. 
In particular, it has been established that the quasipotential depends upon the strength of the interaction, in the precise meaning that the stronger the interaction, the lower the quasipotential effective barrier.
An ideal experiment comparing the escape times of a JJ subject only to noise with those of a junction subject to the same noise and an axion field could  reveal the presence of axions. This would be evidenced by a decrease in the mean escape time, and the magnitude of this decrease would provide a quantitative estimate of the JJ-axion interaction.
Furthermore, the decrease should persist at any noise value, even if very small, i.e., as small as necessary to achieve the desired SNR. Finally, numerical simulations demonstrated that the observed behavior holds for different bias points, thus providing an additional tunable parameter for an experimental setup.

From a practical point of view, a possible detection scheme can follow two different strategies, in line with Ref.~\cite{Grimaudo22}. In the first scheme the energy ratio $\varepsilon$ of a single JJ is tuned to sweep a considerable range of values, from $\varepsilon\ll1$, where the JJ does not feel the presence of axion, up to the resonance condition $\varepsilon\simeq1$, in which the interaction is favoured. This scheme can be realized adjusting the critical current (e.g., via temperature, electric or magnetic fields), being $\varepsilon\propto1/I_c$. In the second, an array of JJs is used, whose critical currents can be: \emph{i}) \emph{ad-hoc} engineered, so as to have a distribution of many different devices with several $\varepsilon$ values, to look for ``anomalies" in switching times in those JJs that satisfy the $\varepsilon\gtrsim1$ condition; \emph{ii}) as similar as possible, for studying the statistics of switching events. In the latter case, some JJs will switch because of the combined action of thermal noise and bias currents, while other JJs will show different switching times, being perturbed by axions.

A word of caution: the change in potential energy is only one of the ingredients for an accurate calculation of SNR, which requires the estimation of fluctuations for finite sampling, as previously done for Josephson-based single photon detection schemes~\cite{Pied21,Gua21,Piedjou21}, and provides additional information on the sample size needed to determine exclusion graphs or the residual operator characteristic~\cite{Filatrella23}.

A further refinement of this approach could be achieved through the \emph{principle of minimum available noise energy}~\cite{Kautz88,Kautz1994} to determine the quasipotential without stochastic simulations. 

%Shortly, the idea could be to follow the path of the unperturbed JJ to determine the critical point, i.e., the separatrix between the two basins of attraction belonging to different stable points. By doing this for the deterministic noise-free system, it is possible to calculate the energy in the minimum and the minimum work required to bring the system from the initial stable point to this critical point, i.e., exactly the quasipotential. Any other work involving a different trajectory between these two stable points is larger than the actual quasipotential. If the method provides a close estimate of the quasipotential through stochastic lengthy numerical calculations, it may be exploited to search the vast parameter space with straightforward deterministic analysis. 

To conclude, our study aims to provide further insights into the interplay between noise, switching dynamic of the JJ and available signal statistical properties, to enhance the understanding of axion JJ-based detector, and at the same time its robustness and reliability. 
Through the characterization of the noise-induced effects and the understanding of their implications, we wish to contribute to the development of better detectors and of quantum technology devices with improved performances.

%\bibliography{bibliofile}

%%%%%%%%%%%%%%%%%%%%%%%%%%%%%%%%%%%%%%%%%%
\vspace{6pt} 

%%%%%%%%%%%%%%%%%%%%%%%%%%%%%%%%%%%%%%%%%%
%% optional
%\supplementary{The following supporting information can be downloaded at:  \linksupplementary{s1}, Figure S1: title; Table S1: title; Video S1: title.}

% Only for the journal Methods and Protocols:
% If you wish to submit a video article, please do so with any other supplementary material.
% \supplementary{The following supporting information can be downloaded at: \linksupplementary{s1}, Figure S1: title; Table S1: title; Video S1: title. A supporting video article is available at doi: link.}

%%%%%%%%%%%%%%%%%%%%%%%%%%%%%%%%%%%%%%%%%%
\authorcontributions{Conceptualization, R.G. and C.G.; methodology, R.G., D.V., and C.G.; software, R.G.; validation, R.G., D.V. and C.G.; formal analysis, R.G.; investigation, R.G. and C.G.; resources, D.V.; data curation, R.G. and C.G.; writing---original draft preparation, G.F. and C.G.; writing---review and editing, R.G., D.V., B.S., A.T., G.F. and C.G.; visualization, R.G., D.V., B.S., A.T., G.F. and C.G.; supervision, D.V., B.S., G.F. and C.G; funding acquisition, D.V. All authors have read and agreed to the published version of the manuscript.}

\dataavailability{The data presented in this study are available on reasonable request from the corresponding authors.} 

\acknowledgments{R.G. acknowledges financial support from the PRIN Project PRJ-0232 - Impact of Climate Change on the biogeochemistry of Contaminants in the Mediterranean sea (ICCC). All the authors acknowledge the support of the Ministry of University and Research of Italian Government.}

\conflictsofinterest{The authors declare no conflict of interest.} 

%%%%%%%%%%%%%%%%%%%%%%%%%%%%%%%%%%%%%%%%%%
%% Optional
%\sampleavailability{Samples of the compounds ... are available from the authors.}

%% Only for journal Encyclopedia
%\entrylink{The Link to this entry published on the encyclopedia platform.}

\abbreviations{Abbreviations}{
The following abbreviations are used in this manuscript:\\

\noindent 
\begin{tabular}{@{}ll}
JJ & Josephson junction\\
RCSJ & resistively and capacitively shunted junction
\end{tabular}
}

%%%%%%%%%%%%%%%%%%%%%%%%%%%%%%%%%%%%%%%%%%

%%%%%%%%%%%%%%%%%%%%%%%%%%%%%%%%%%%%%%%%%%
\begin{adjustwidth}{-\extralength}{0cm}
%\printendnotes[custom] % Un-comment to print a list of endnotes

%\bibliography{bibliofile}

\PublishersNote{}
\end{adjustwidth}
\end{document}